\documentstyle[12pt]{article}
\begin{document}
\input psfig
\begin{titlepage}
\begin{center}
\today     \hfill    WM-98-106\\

\vskip .25in

{\large \bf  Photoproduction of of $\ell=1$ Baryons: \\
Quark Model versus Large N$_c$}

\vskip 0.3in

Carl E. Carlson and Christopher D. Carone 

\vskip 0.1in

{\em Nuclear and Particle Theory Group \\
     Department of Physics \\
     College of William and Mary \\
     Williamsburg, VA 23187-8795}


        
\end{center}

\vskip .1in

\begin{abstract}
We consider the electromagnetic decays of the orbitally-excited SU($6$) 
{\bf 70}-plet baryons, to compare the predictions of the naive quark model 
with those of large-$N_c$ QCD. The helicity amplitudes measured in $N^*$ 
photoproduction are computed in a large-$N_c$ effective field theory, based 
on a Hartree approximation, and the amplitudes are fit to the current 
experimental data. Our results indicate that the success of the naive quark 
model predictions cannot be explained by large-$N_c$ arguments alone.  This 
is consistent with the conclusions of an earlier study of 
the $N^*\rightarrow N\pi$ decays, that utilized the same approach.
\end{abstract}

\end{titlepage}

\newpage
\renewcommand{\thepage}{\arabic{page}}
\setcounter{page}{1}
\section{Introduction} \label{sec:intro} \setcounter{equation}{0}

That the constituent quark model can work remains something of a mystery.  
A simple realization of the model treats baryon states as spin-flavor 
SU($6$) wave functions and treats interactions that lead to the decay or 
production of the states as single-quark operators.  A group theoretical 
strategy can be adopted in connection with this model, so that instead of 
calculating specific dynamics, one just finds the structure of the hopefully 
few allowed one-body operators that can contribute, and then fits the 
coefficients of these operators to a subset or all of the data and attempts 
to make predictions.  One might think that there would be large corrections 
to this type of model, for example via multiquark strong interactions that 
could not simply be absorbed into the binding potential.

Recently, it has been noted that large-$N_c$ quantum chromodynamics (QCD) 
can be used to show that corrections to the simplest operators are in a 
number of cases suppressed by powers of $1/N_c$.  These cases include 
analyses of operators that contribute to the axial current matrix elements 
and the masses of the SU($6$) ground state baryons, and to operators that 
would modify the SU($6$) wave functions even for low spin 
states \cite{cgo,ml,djm1,djm2}.  

However, a different large-$N_c$ result has been noted in the decays of the
negative parity P-wave baryons, the {\bf 70}-plet of SU($6$), into ground 
state baryons plus pions.  Here one may write down all the independent 
one-quark and multiquark operators that are allowed \cite{cgkm}.  One finds 
that, while some of the two-body operators are suppressed by a power 
of $1/N_c$ for low spin states, a well-defined subset of the two-body 
operators are unsuppressed relative to the one-body operators in the 
large $N_c$ limit.  The reason for the lack of suppression is clear.  The 
two-body operators themselves contain a suppresion by a factor of $1/N_c$, 
but coherent effects in the matrix elements can give a compensatory factor 
of $N_c$.  There is still something of a mystery, however.  An explicit 
numerical analysis of the decays of the {\bf 70}-plet into a pion plus a 
member of the {\bf 56}-plet, the ground state, showed that the coefficients 
of the two-body operators are small.

In this paper,  we shall investigate the radiative decays of 
the {\bf 70}-plet, or equivalently their radiative production.  In 
particular we will be interested in the guidance that we can get from 
large-$N_c$ QCD regarding which operators give the leading contributions 
to the decay matrix elements.

The study of the radiative {\bf 70}-plet decays using SU($6$) wave functions 
and just one-body operators is long since textbook material \cite{close}.  
There are 3 one-body operators, and hence 3 coefficients that may be used to 
describe the 19 independent decay amplitudes of the {\bf 70}-plet into 
nucleons. The simple description, as we shall review below, works reasonably 
well for most decays. 

One might expect sizable corrections from two-body operators to the simple
picture.  Again, two-body operators are operators where a second quark is
interacting with the quark that absorbs the photon, and doing so in ways that
cannot be  just incorporated into the binding interaction.  The two-body 
operators involve strong interaction corrections to a simple photon 
interaction, and are not in a regime where one would expect the corrections 
to be perturbative.  They could be big.

In terms of the large-$N_c$ limit, since only one quark is excited in the
{\bf 70}-plet, it is elementary to show that matrix elements of the one-body
operators are of order 1 in the $N_c$ expansion.  Now consider the matrix
elements of the two-body operators.  The two-quark operators themselves are of
order $1/N_c$, including factors of the QCD coupling in with the operator.
However, because of the many possible choices for the second quark when a
matrix element is taken, there is a possible $O(N_c)$ coherence factor.  This 
leads to the result that certain two-quark operators are also overall $O(1)$ 
in the large $N_c$ expansion.  So in any analysis of the {\bf 70}-plet decays 
it would seem an oversight to exclude the two-body operators.  We should test 
the significance of the two-quark operators by including them in fits to 
data, and see if their presence can be shown.  (We shall find that the 
reverse is true.)

Before describing our work in more detail, let us make two comments.  One is
that some of the two-body operator matrix elements are indeed suppressed by
$1/N_c$.  This can happen if they are proportional to quark spin, in which case
they add up to the total quark spin of the state,  which for low spin states
does not compensate for the $O(1/N_c)$ size of the operator before the matrix
element is taken.  The other is that spin-spin interaction effects can, for
high spin states, build up and significantly modify the SU($2f$) form of the
wave function.  The effect is small for low spin states.   Hence if the
$\Delta$ can be treated as a low spin state, which a spin-3/2 state surely is
in the context of large $N_c$, the SU($2f$) wave functions should be good for
it.  Extrapolating to $N_c = 3$, we shall treat the $\Delta$ as a good SU($6$)
state, but shall also keep in mind that this may be less well justified for
the $\Delta$ than for the nucleon.

In Section~\ref{sectiontwo} we explain the applications of the high $N_c$ limit
to radiative decays of the excited baryons and give other information about our
calculation, Section~\ref{sectionthree} gives our numerical results, and
Section~\ref{sectionfour} gives our conclusions.

\section{Framework}  \label{sectiontwo}

{\em Formalism.} Our basic approach is the same as that of 
Refs.~\cite{cgo,cgkm}.  We assume that we can represent the large-$N_c$ 
baryon states in the tensor product space of the spin and flavor indices of 
the $N_c$ valence quarks.  Our states therefore have the same spin, flavor, 
and orbital angular momentum structure as representations of nonrelativistic 
SU($6$)$\times$O($3$).  If we were to consider the limit where the quarks 
are heavy compared to $\Lambda_{QCD}$, this assumption clearly would be 
correct, since the nonrelativistic quark model description of the baryon 
states can be then derived from QCD.  The spin-flavor states with the same 
spatial wave functions are degenerate in the limit of large quark masses, and 
form complete SU($6$)$\times$O($3$) multiplets.   Although it appears that 
this description would break down completely for baryons containing quarks 
that are much lighter than $\Lambda_{QCD}$, one must keep in mind 
that the splittings between states of low spin within a given multiplet are 
not only suppressed by $1/m_q$, where $m_q$ is the quark mass, but also by 
$1/N_c$. Thus, in the large-$N_c$ limit, we expect that the states with 
lowest spin within each spin-flavor multiplet should be well described in 
the same tensor product space that is appropriate for large $m_q$.  We only 
need assume that there is no discontinuity in the description of the baryon 
states as we gradually lower the quark mass.

The approximate symmetry described above is an unusual one, in that overall 
it is badly broken within every multiplet.  However, since the dimension of 
each multiplet grows with $N_c$, while the symmetry breaking effects scale 
as $1/N_c$, it is at least possible to make reliable predictions at the 
low-spin end of each multiplet.  As we describe below, we will ultimately 
work with $N_c=3$ baryon wave functions, where the distinction between 
the ``top" and ``bottom" of a multiplet is not completely clear.  To avoid 
any ambiguity, we choose to work with complete spin-flavor representations, 
keeping in mind that our results may be less reliable for the states 
of highest spin.  

Assuming this basis, we can write the Hartree-Fock interaction Hamiltonian 
as
\[
H_{int}=\sum_{n=1}^{N_c} \, \sum_{\{\alpha_1,\ldots,\alpha_n\} \subset 
\{1,\ldots,N_c\}} \int d^3 r_{\alpha_1} \times \cdots \times d^3 r_{\alpha_n} 
\Phi(r_{\alpha_1})_{\alpha_1}^\dagger 
\otimes
\]
\begin{equation}
\cdots \otimes \Phi(r_{\alpha_n})_{\alpha_n}^\dagger 
{\cal O}(r_{\alpha_1},\ldots,r_{\alpha_n})
\Psi(r_{\alpha_1})_{\alpha_1}\otimes \cdots \otimes 
\Psi(r_{\alpha_{n-1}})_{\alpha_{n-1}}
\otimes \Psi_*(r_{\alpha_n})_{\alpha_n} \,\,\, ,
\label{eq:ham}
\end{equation}
where ${\cal O}$ is any quark operator of interest. The $\Phi$ and $\Psi$ 
factors are the individual quark wave functions in the {\bf 56} and 
{\bf 70}-plet baryons, respectively.  The asterisk on the $n^{th}$ $\Psi$ 
wave function indicates that it is the one corresponding to the 
orbitally-excited quark.  Notice that Eq.~(\ref{eq:ham}) sums over all 
possible multiquark interactions involving $n$ quarks, and then over 
all possible values of $n$.  By Witten's power counting arguments, we expect a 
general $n$-body interaction to scale as $1/N_c^{n-1}$ \cite{witten}.  Since a 
transition between the {\bf 70} and {\bf 56}-plet baryons always involves the 
deexcitation of the orbitally-excited quark, the lowest-order operators will 
be one-body, acting only on the excited quark line. Higher-body operators 
will involve the orbitally-excited quark as well as one or more of the other 
valence quarks, and be suppressed by powers of $1/N_c$.  

The most useful way of thinking about the interaction Hamiltonian in
Eq.~(\ref{eq:ham}) is in terms of its spin and flavor transformation 
properties. Assuming $f$ quark flavors, each large-$N_c$ baryon state can 
be viewed as a $(2f)^{N_c}$ dimensional vector, while $H_{int}$ can be 
though of a $(2f)^{N_c}\times(2f)^{N_c}$ matrix acting in this space.  
While we cannot explicitly evaluate this matrix (by computing the Hartree 
wave functions and performing the integration in Eq.~(\ref{eq:ham})) we can 
determine the form of $H$, up to unknown coefficients, from symmetry 
considerations.   The argument goes as follows:

Since the wave functions $\Phi$ and $\Psi$ are computed from the portion of the
Hartree potential that is zeroth order in spin and flavor symmetry breaking, 
these wave functions are spin-flavor independent and spherically symmetric. 
Thus, they are proportional to identity matrices in the spin-flavor space, and
can be replaced by ordinary functions
\begin{equation}
\Phi({\bf r}) \rightarrow \phi(r), \,\,\, \Psi({\bf r}) \rightarrow \psi(r) 
= \phi(r) \,\,\, ,
\end{equation}
where $r=|{\bf r}|$ is the radial coordinate.  The last equality follows
from the observation that the Hartree potential generated by order $N_c$
quarks should remain unaffected by the excitation of a single quark.  
The wave function of the excited quark, on the other hand, has the form
\begin{equation}
\Psi_*({\bf r})=f(r) Y_{l=1,m}(\theta,\varphi)=
\sqrt{\frac{3}{4\pi}} f(r)({\bf \hat{r}}\cdot {\bf \varepsilon}_m)
\end{equation}
where $f(r)$ is a spherically-symmetric function, and $Y$ is a spherical 
harmonic; the $\ell=1$ polarization vectors $\varepsilon_m$ are given by
\begin{equation}
\varepsilon_1=\frac{1}{\sqrt{2}}\left(\begin{array}{c} -1 \\ -i \\ 0 
\end{array}\right) , \, 
\varepsilon_0=\left(\begin{array}{c} 0 \\ 0 \\ 1 
\end{array}\right) , \, 
\varepsilon_{-1}=\frac{1}{\sqrt{2}}\left(\begin{array}{c} 1 \\ -i \\ 0 
\end{array}\right) \, . 
\end{equation}
The interaction Hamiltonian is therefore an integral of the operator ${\cal O}$
times $2N_c$ spherically symmetric, spin-flavor independent functions, times
${\bf r}\cdot {\bf \varepsilon}_m$.   Thus, we can evaluate an  $n$-body 
contribution to $H$ by first determining the spin-flavor transformation 
properties of the QCD operator ${\cal O}$ under the 
SU($2f$)$^{N_c}\times $SU($2f$)$^{N_c}$ spin-flavor symmetry of the 
zeroth-order part of Hartree Hamiltonian.  Notice that this corresponds to 
a different SU($2f$) symmetry for every quark line. We then write down the 
most general set of operators transforming in the same way, out of the spin 
and flavor generators, $\sigma^a$ and $\lambda^a$, the polarization vector 
$\varepsilon_m$, and the momentum ${\bf k}$ of the light decay product.  We 
discuss the construction of these operators in greater detail below.  We may 
evaluate any matrix elements of interest by letting the operators 
act on the tensor representations of the baryon states; the undetermined
coefficients can then be fit to the corresponding experimental data.

What makes this approach nontrivial is that certain types of operators that 
are subleading in $1/N_c$ may nonetheless have matrix elements that are as 
important as those of the leading operators.  For example, the matrix element 
of a two-body operator that contributes to a {\bf 70}-plet baryon decay 
necessarily involves a spin-flavor matrix acting on the excited quark line, 
as well as a spin-flavor matrix acting on a nonexcited quark line, summed 
over the $N_c-1$ ground state quarks.  If this sum is coherent, it will 
scale as $N_c$, and the matrix element of this two-body operator will be of 
the same order as those of the leading one-body operators.  Thus, by 
determining what types of sums are potentially coherent on the low-spin 
baryon states, we can restrict the set of operators that appear in the 
low-energy effective theory. Since the one-body operators are often 
identified with the naive quark model predictions, our large $N_c$ arguments 
imply that there are additional operators, with distinct spin-flavor 
transformation properties, that may be of equal importance in describing 
the experimental data.   A fit to the data will then determine whether this 
is actually the case.   The different spin-flavor matrices and their 
properties were summarized in Ref.~\cite{cgkm}, and we provide them again 
for convenience:

$\bullet$ Constant terms -- these always add coherently when summed over quark 
lines, but have no nontrivial spin-flavor structure.  Thus, these terms are 
irrelevant.

$\bullet$ Spin terms - These involve the sum
\begin{equation}
\sum_{{\rm quarks} \,\,\,\, \alpha} \sigma_\alpha^j
\end{equation}
where $\sigma^j$ is a pauli matrix.  These are of order one on
the low spin states \cite{cgo}, and are thus associated with a factor 
of $1/N_c$ relative to the leading terms.

$\bullet$ Flavor terms --  These involve the sum
\begin{equation}
\sum_{{\rm quarks} \,\,\,\, \alpha} \lambda_\alpha^a
\end{equation}
where $\lambda^a$ is a Gell-Mann matrix, for flavor SU($3$).  These terms
can add coherently, depending on the flavor quantum numbers of the
large $N_c$ baryon state.

Below, we will use this approach to write down the operators relevant in 
computing the $N\rightarrow N^*$ helicity amplitudes.  After we have 
determined the set of operators whose matrix elements are potentially of 
leading order, we will evaluate the matrix elements using the baryon wave 
functions corresponding to $N_c=3$.  This allows us to avoid any ambiguity in 
determining which states in a large-$N_c$ baryon multiplet should be 
identified with the physical states observed in nature.  Note that this 
feature makes our approach different from Refs.~\cite{djm1,djm2,py}, 
where a reasonable embedding of the physical baryon states within the 
large-$N_c$ multiplet is assumed.

{\em Operators.}  Since we are interested in the $N^*N\gamma$ coupling, we 
have at the quark level the QED vertex,
\begin{equation}
\overline{q} Q \gamma^\mu q A_\mu
\end{equation}
where the charge $Q$ is a matrix in SU($3$) flavor space, 
$Q=$diag$(2/3,-1/3,-1/3)$. Thus, at lowest order in flavor symmetry 
breaking, our operators will involve one factor of the matrix $Q$, acting on 
the flavor indices of the baryon states.   Other quark-level operators arise 
when the operator above acts on one quark line, with gluon exchange involving 
other lines. From the start, we will work in Coulomb gauge, where $A^0=0$ and 
$\nabla \cdot {\bf A} = 0$. Although gauge invariance will not be manifest, 
this choice is physically equivalent to any other, and greatly simplifies 
the operator analysis.  

The set of interactions that we consider in this paper include the one-body
operators
\begin{equation}
a_1 Q_* \vec{\varepsilon}_m \cdot \vec{A} \,\,\, ,
\end{equation}
\begin{equation}
i b_1 Q_* \vec{\varepsilon}_m \cdot \vec{\nabla} 
(\vec{\sigma}_*\cdot\vec{\nabla}
\times \vec{A}) \,\,\, ,
\end{equation}
\begin{equation}
i b_2 Q_* \vec{\sigma}_*\cdot \vec{\nabla}(\vec{\varepsilon}_m 
\cdot \vec{\nabla}
\times \vec{A}) \,\,\, , 
\end{equation}
and the two-body operators,
\begin{equation}
i c_1  (\sum_{\alpha \neq *}Q_\alpha)( \vec{\sigma}_* \times 
\vec{\varepsilon}_m \cdot 
\vec{A})
\end{equation}
\begin{equation}
c_2 (\sum_{\alpha \neq *} Q_\alpha \vec{\sigma}_\alpha) 
\cdot \vec{\varepsilon}_m (\vec{\sigma}_*
\cdot \vec{A})
\end{equation}
\begin{equation}
c_3 (\sum_{\alpha \neq *} Q_\alpha \vec{\sigma}_\alpha )\cdot \vec{\sigma}_* 
(\vec{\varepsilon}_m \cdot \vec{A})
\end{equation}
\begin{equation}
c_4 (\vec{\sigma}_*\cdot \vec{\varepsilon}_m) (\sum_{\alpha \neq *} Q_\alpha 
\vec{\sigma}_x) \cdot \vec{A}
\end{equation}
\begin{equation}
i d_1  (\sum_{\alpha \neq *}Q_\alpha) \vec{\varepsilon}_m \cdot \vec{\nabla} 
(\vec{\sigma}_* \cdot \vec{\nabla} \times \vec{A})
\end{equation}
\begin{equation}
d_2 (\sum_{\alpha \neq *} Q_\alpha \vec{\sigma}_\alpha) \cdot \vec{\nabla} 
(\vec{\varepsilon}_m \cdot \vec{\nabla})( \vec{\sigma}_* \cdot \vec{A})
\end{equation}
\begin{equation}
d_3 (\sum_{\alpha \neq *} Q_\alpha \vec{\sigma}_\alpha) \cdot \vec{\nabla} 
(\vec{\sigma}_* \cdot
\vec{\nabla}) (\vec{\varepsilon}_m \cdot \vec{A})
\end{equation}
\begin{equation}
d_4 \left[(\sum_{\alpha \neq *} Q_\alpha \vec{\sigma}_\alpha) 
\times \vec{\sigma}_* \cdot
\vec{\nabla}\right] (\vec{\varepsilon}_m \cdot \vec{\nabla}
\times \vec{A}) \,\,\, .
\end{equation}
\begin{figure}[ht]
\centerline{\psfig{file=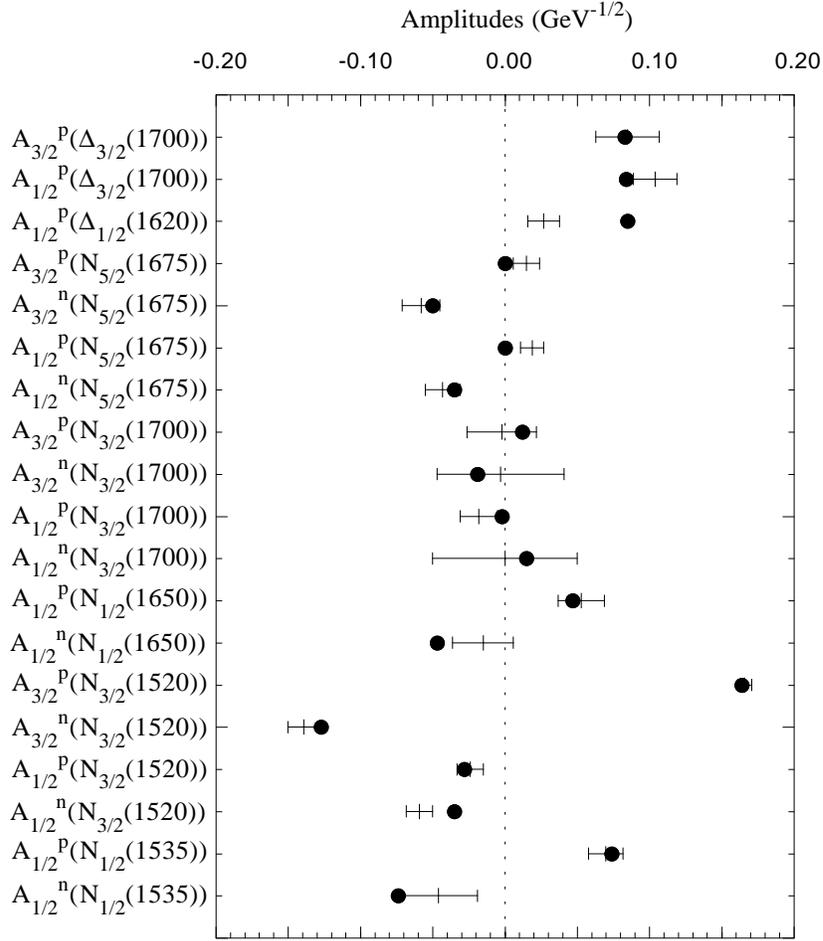,width=11cm}}
\caption{Results of a fit to the one-body operators and mixing angles. The fit 
corresponds to the parameter set $a_1=0.615$, $b_1=-0.294$, $b_2=-0.297$, 
$\theta_{N1}=0.597$ and $\theta_{N3}=3.060$.  The resulting numerical values
of the helicity amplitudes are also given in Table~\ref{table2}.}
\label{figure1}
\end{figure}
where the asterisk denotes the excited quark line. We assume that the 
derivatives in these operators are suppressed by the scale $\Lambda_{QCD}$, 
which we have left implicit, for notational convenience.  The fact that we 
find only three linearly independent one-body operators is consistent with 
a long-known quark model result.  For example, in Ref.~\cite{close} the 
helicity amplitudes that we consider are computed by assuming the photon 
couples to the current $J$, where $J_+=A L_+ + B S_+ + C S_z L_+$.  
Here $S_\pm$ and $L_\pm$ represent the quark spin and orbital angular 
momentum raising and lowering operators, while $A$, $B$ and $C$ 
are SU($6$)-invariant coefficients.  These are related to the coefficients of 
our one-body operators above by $a_1=\sqrt{3/2}A$, $b_1=-\sqrt{3/2}B$, and 
$b_2 = -\sqrt{3/2}C$\footnote{See the comment after Eq.~(\ref{eq:tpt}), 
regarding the factors of momentum in our operators.}.  As 
we describe below, the other operators we present 
completely span the space of possible two-body interactions.  A three-body 
operator would involve two sums over the $\alpha \neq *$ quark lines; at 
lowest order in $Q$, at least one of these sums would be of the form 
$\sum \sigma_\alpha$, which is subleading in $1/N_c$, by the rules 
described earlier in this section.   Note that we can generate additional 
operators if we include pions in our theory, since the pion-quark-quark 
coupling involves the spin-flavor structure $\sigma^j\lambda^a$, and can 
yield a coherent effect when summed over $N_c-1$ quark lines. However, in 
our framework such couplings match onto pion-baryon interactions in the 
low-energy effective theory, and thus any new spin-flavor operators arise 
as only as loop effects, which should be small.  This interpretation is 
consistent with our numerical results, which indicate that our operator 
list above is sufficient to give an excellent description of the data.    

The helicity amplitudes $A_{1/2}$ and $A_{3/2}$ which we would like to study 
are defined via the photoproduction differential cross section of the $N^*$ 
baryons.  There are $19$ independent helicity amplitudes (not connected by 
isospin) for {\bf 70}-plet production off $p$ or $n$ targets. If on the 
other hand we would like to think about the problem in the $N^*$ rest frame, 
the $N^*\rightarrow N \gamma$ partial decay widths are proportional to the 
sum of these squared amplitudes.  For a complete discussion we refer the 
reader to Ref.~\cite{rpp}, and references therein.  The definitions that we 
require for our analysis are
\begin{equation}
A_{1/2} = K \, \xi \, \langle N^*, \,\, s_z=\frac{1}{2} ; \gamma ,\,\, 
\varepsilon_{+1} | H_{int}| N,\,\, s_z=-\frac{1}{2}\rangle
\label{eq:a12}
\end{equation}
\begin{equation}
A_{3/2} = K \, \xi \, \langle N^*, \,\, s_z=\frac{3}{2} ; \gamma ,\,\, 
\varepsilon_{+1} | H_{int} | N,\,\, s_z=\frac{1}{2} \rangle  \,\,\, ,
\label{eq:a32}
\end{equation}
where $s_z$ is the $z$ component spin in the $N^*$ rest frame, and the 
$\varepsilon_m$ shown indicates the photon polarization.  $K$ is a kinematical
factor, given by $[4\pi\alpha m^2_{N^*}/(m_{N^*} - m^2_N)]^{1/2}$, and the
states have the standard nonrelativistic normalization.  The 
factor $\xi$ is the sign of the $N^* \rightarrow N \pi$ amplitude by
which the $N^*$ is detected in photoproduction experiments. This takes
into account the proper sign convention that experimenters use when they
extract the helicity amplitudes from their data, and also renders our
results independent of the sign conventions of the states.  We evaluate the 
matrix elements in Eqs.~(\ref{eq:a12}) and (\ref{eq:a32}) by allowing the our 
one- and two-body operators to act on tensor representations of the baryon 
states.  It is important to point out that we have restricted our set of 
two-body interactions to include only those operators whose matrix elements 
in (\ref{eq:a12}) and (\ref{eq:a32}) are linearly independent over the set 
of nonstrange {\bf 70}-plet states.  In this subspace, we find that the 
combination of our large-$N_c$ power counting rules, and the constraint of 
conservation of angular momentum restricts the number of two-body 
interactions to 11; we then found 3 additional, nontrivial operator relations 
that reduced this set to the 8 operators shown above\footnote{Our original 
set of 11 two-body operators included  $c_5 \, (\sum_{\alpha \neq *}Q_\alpha) 
\vec{\varepsilon}_m \cdot \vec{A}$, $c_6 \, (\sum_{\alpha \neq *}Q_\alpha 
\vec{\sigma}_\alpha) \times \vec{\varepsilon}_m \cdot \vec{A}$, and 
$i d_5 \, \vec{\varepsilon}_m \cdot \vec{\nabla} (\sum_{\alpha \neq *} 
Q_\alpha \vec{\sigma}_\alpha)\cdot \vec{\nabla}\times \vec{A}$. However we 
found that the matrix elements of these operators were related to those 
given in the text by $c_5=-a_1$, $c_6=-b_1+b_2$ and $d_5=-b_1$.}. The actual 
computation was done using code written in MAPLE\footnote{The matrix elements
of the one-body operators are easy to do by hand; the code is helpful in 
evaluating the two-body operators.}, and fits to the data were performed 
using standard FORTRAN minimizations routines.  

\begin{figure}[ht]
\centerline{\psfig{file=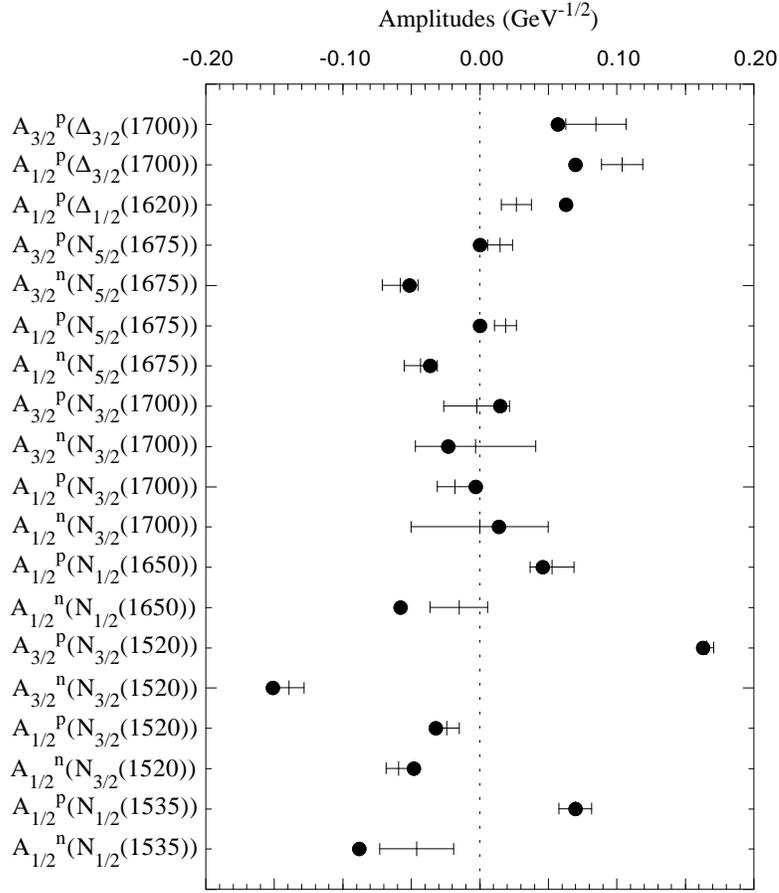,width=11cm}}
\caption{Results of the fit including the one-body operators and operator 
$c_3$, with mixing angles fixed.  The fit corresponds to the parameter set 
$a_1=0.816$, $b_1=-0.299$, $b_2=-0.308$, $c_3=-0.072$, $\theta_{N1}=0.610$ and 
$\theta_{N3}=3.040$; see also Table~\ref{table3}.} 
\label{figure2}
\end{figure}
\section{Results} \label{sectionthree}
\setcounter{equation}{0}

In this section we present our fits to the helicity amplitudes $A_{1/2}$ and
$A_{3/2}$ of the nonstrange $\ell=1$ baryons.  We take our experimental data 
points and error bars from the 1996 Review of Particle Properties \cite{rpp}.
Our possible set of free parameters includes the coefficients  of the one- 
and two-body operators, defined in the previous section, as well as two 
mixing angles: $\theta_{N1}$ and $\theta_{N3}$.  These mixing angles are 
necessary to specify the spin-1/2 and spin-3/2 nucleon mass eigenstates,
\begin{equation}
\left[\begin{array}{c} N(1535) \\ N(1650) \end{array} \right] =
\left[\begin{array}{cc}  \cos\theta_{N1} & \sin\theta_{N1} \\
                       -\sin\theta_{N1} & \cos\theta_{N1} \end{array}\right] 
\left[\begin{array}{c} N_{11} \\ N_{31}\end{array} \right]
\end{equation}
and
\begin{equation}
\left[\begin{array}{c} N(1520) \\ N(1700) \end{array} \right] =
\left[\begin{array}{cc}  \cos\theta_{N3} & \sin\theta_{N3} \\
                       -\sin\theta_{N3} & \cos\theta_{N3} \end{array}\right] 
\left[\begin{array}{c} N_{13} \\ N_{33}\end{array} \right]  \,\,\, , 
\label{eq:tpt}
\end{equation}
where $N_{ij}$ represents a state with total quark spin $i/2$ and total
angular momentum $j/2$.  Note that we absorb any factors of 
momentum/$\Lambda_{QCD}$ that appear in the operators into our definition of 
the fit coefficients.  This is consistent since we are extracting the leading 
order contributions to the amplitudes in $1/N_c$, and in this limit each 
SU(6) multiplet is not split.  We do take into account the actual mass 
spectrum of the states when evaluating the kinematical factor $K$.  We will 
first show that a fit retaining only the one-body operators, and allowing the 
mixing angles to vary, provides a reasonable (though by no means perfect) 
description of the experimental data.  We will find that the most likely 
values of the mixing angles obtained in this fit are in excellent agreement 
with the results of Ref.~\cite{cgkm}, which were obtained by a completely 
independent analysis of the $N^*\rightarrow N \pi$ decays.   We will then 
show that inclusion of the two-body operators does not significantly improve 
the fit, so that our conclusions are qualitatively similar to those described 
in Ref.~\cite{cgkm}.  Finally, we will show that we can achieve a drastic
improvement in the fit involving the one-body operators alone by 
discarding a single experimental data point.  The improvement is so
dramatic, we are left with the suspicion that one of the helicity amplitudes 
is affected by physics that we do not include in our analysis.

\begin{figure}[ht]
\centerline{\psfig{file=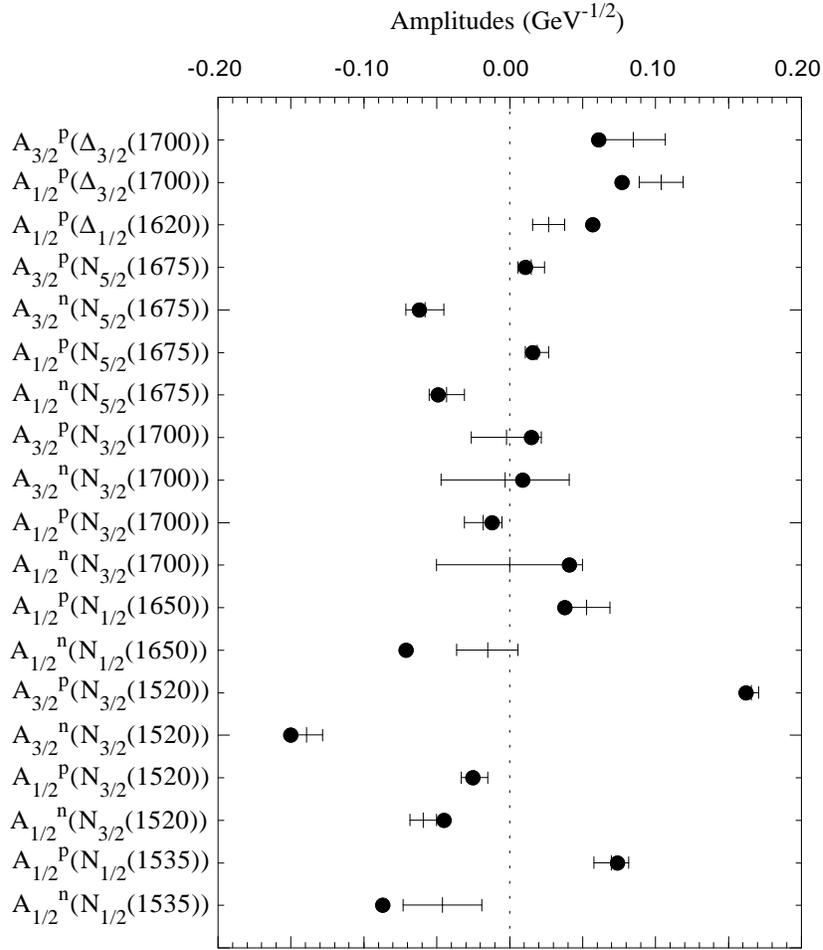,width=11cm}}
\caption{Results of the fit including all the one- and two-body operators 
given in Section~2, with mixing angles fixed.  The fit corresponds to the 
parameter set $a_1=0.808$, $b_1=-0.192$, $b_2=-0.498$,$c_1=-0.087$,
$c_2=-0.036$, $c_3=-0.077$, $c_4=0.049$, $d_1=0.077$, $d_2=0.009$, 
$d_3=0.013$, $d_4=0.012$, $\theta_{N1}=0.610$ and $\theta_{N3}=3.040$; see 
also Table~\ref{table4}.} 
\label{figure3}
\end{figure}
Fig.~1 shows the fit obtained by retaining the three one-body operators, 
allowing their coefficients as well as the mixing angles $\theta_{N1}$ 
and $\theta_{N3}$ to vary. The $\chi^2$ for this fit is $52.9$, with
\begin{figure}[ht]
\centerline{\psfig{file=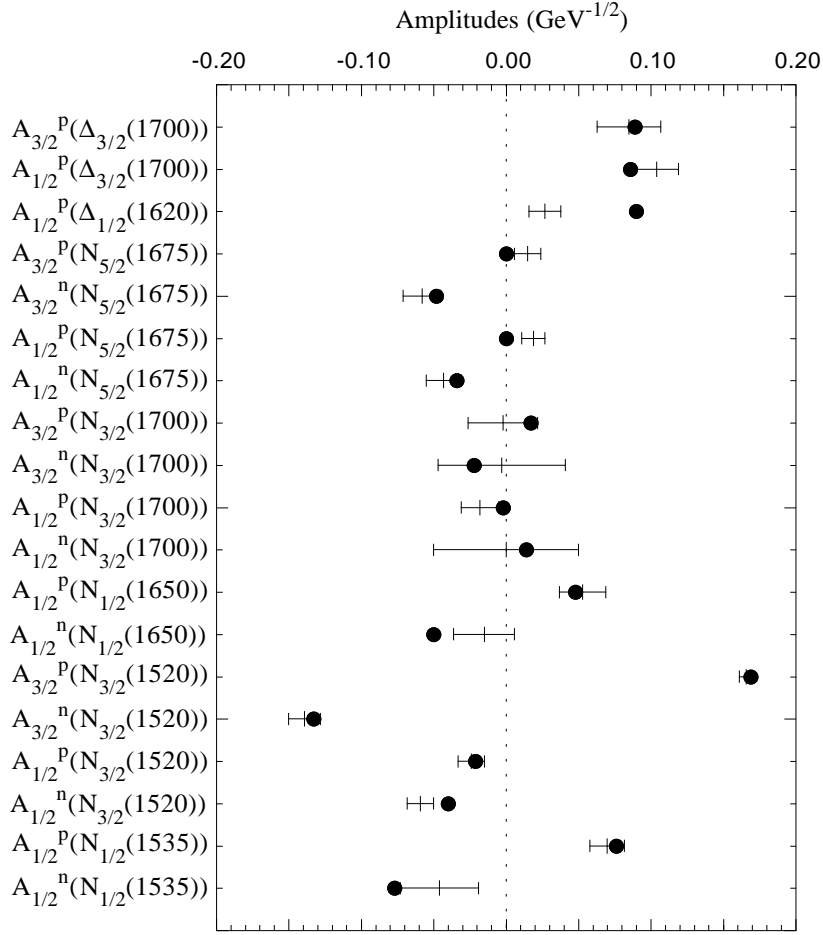,width=11cm}}
\caption{Results of the fit to the one-body operators and mixing angles, with 
$A_{1/2}(\Delta^+(1620))$ omitted from the $\chi^2$.  The fit corresponds
to the parameter set $a_1=0.642$, $b_1=-0.268$, $b_2=-0.297$, 
$\theta_{N1}=0.591$ and $\theta_{N3}=3.031$; see also Table~\ref{table5}.} 
\label{figure4}
\end{figure} 
the greatest contributions coming from the amplitudes 
$A_{1/2}(\Delta^+(1620))$ ($\Delta \chi^2 = 27.5$) and 
$A_{1/2}(N^0(1520))$ ($\Delta \chi^2=7.7$).  What is perhaps most
striking about this fit is the favored values of the mixing
angles, $\theta_{N1}=0.60 \pm 0.16$ and $\theta_{N3}=3.06\pm 0.11$.
In Ref.~\cite{cgkm} the same mixing angles were extracted from
a large-$N_c$ analysis of $N^*\rightarrow N \pi$ decays, and the
values $\theta_{N1}=0.61\pm 0.09$ and $\theta_{N3}=3.04\pm 0.15$
were obtained.  Thus, the two analyses are in complete agreement.
(Furthermore, both are consistent with the mixing angles 
found in earlier studies of the decays of baryon resonances that
are not in the context of large-$N_c$ \cite{hey}.) Therefore, it is 
justified to fix the mixing angles at the values extracted
from the fit to the pion decays, and explore how the helicity amplitude
fit improves as we introduce the two-body operators.

We can begin exploring the effects of the two-body operators by
introducing them one at a time.  To be precise, we now do a series
of fits in which we include the three one-body operators as well as
a single two-body operator, with the mixing angles fixed at their preferred
values.   The change in $\chi^2$ gives some indication of the impact of the 
given two-body operator on the quality of the fit, or equivalently, on the 
two problematic amplitudes $A_{1/2}(\Delta^+(1620))$ and $A_{1/2}(N^0(1520))$.
This approach can be misleading if the inclusion of more than one two-body 
operator leads to near cancellations between their matrix elements in some of 
the helicity amplitudes, but not in others.  Such a situation can arise 
accidentally, as a consequence of the particular choice of operator basis.  
However, we will also consider a fit that includes all the one- and two-body
operators, and verify that our conclusions remain unchanged.

\begin{table}
\begin{center}
\begin{tabular}{ccccccccc} \hline\hline
Two-body Operator   &  $c_1$ & $c_2$ & $c_3$ & $c_4$ & $d_1$ & $d_2$ 
& $d_3$ & $d_4$ \\ \hline
$\chi^2$ &  52.85 &  52.04 &  39.21 &  52.91 &  48.38 &  52.81 &  52.23 &  
52.59 \\ \hline \hline
\end{tabular}
\caption{$\chi^2$ for fits including the three one-body operators,
and a single two-body operator.  Mixing angles are fixed at 
$\theta_{N1}=0.61$ and $\theta_{N3}=3.04$, discussed in the text. For 
comparison, the fit with the one-body operators alone had $\chi^2=52.97$.}
\label{table1}
\end{center}
\end{table}

The resulting $\chi^2$ values for fits including the three one-body operators
and a single two-body operator are shown in Table~1.  The only
operator whose inclusion had a non-negligible impact on the quality of
the fit was operator $c_3$; the corresponding predictions for the helicity 
amplitudes are displayed in Fig.~2.  While there is a slight improvement in 
the agreement between theory and data, the two problematic amplitudes still 
remain.  

A fit that includes all the one- and two-body operators is shown in Fig.~3. 
Although we have added 8 operators in addition to those of the fit in 
Fig.~1, the $\chi^2$ only improves to $28.87$, with the greatest contributions 
coming from the amplitudes $A_{1/2}(\Delta^+(1620))$ ($\Delta \chi^2= 7.39$) 
and $A_{1/2}(N^0(1650))$ ($\Delta \chi^2=7.13$).  Thus, the $\chi^2$ per 
degree of freedom in our purely one-body fit, $3.8$, reduces to $3.6$ when 
we include all of the remaining operators.  While we might have guessed, 
given the arguments in Section~2, that at least some of the two-body 
operators would have a significant impact on the goodness of the fit, what 
we have found is that none of them play a crucial role in describing the 
experimental data.

The fit in Fig.~4 involves the same assumptions as in Fig.~1, except that
we now omit the contribution of $A_{1/2}(\Delta^+(1620))$ to the total 
$\chi^2$. The resulting fit has $\chi^2=21.45$, or 1.65 per degree of 
freedom.  If we allow for the possibility that there is something wrong with 
this experimental datum, then one can argue that the result from the one-body 
fit alone is good enough to obviate consideration of the higher-body operators.

\section{Conclusions} \label{sectionfour}

What we have found by studying the electromagnetic couplings of the 
$\ell=1$ baryons are results that are qualitatively similar to those found in 
the analysis of the $N^*\rightarrow N\pi$ decays in Ref.~\cite{cgkm}.  The 
one-body interactions appear to give a good description of the observed 
phenomenology, better than one would naively expect from large-$N_c$ 
arguments presented in Section~2. As the authors of Ref.~\cite{cgkm} pointed 
out, this does not necessarily imply that these large-$N_c$ arguments are 
wrong, but could indicate that the two-body operators in the example we've 
considered have small coefficients by chance.  The difficulty with this 
explanation is that it seems less probable, given the total number of 
two-body operators that have been evaluated in Ref.~\cite{cgkm} and here.  
Thus, we conclude in the same way as Ref.~\cite{cgkm}, by noting that 
perhaps there is something more to the success of the naive quark model 
than large-$N_c$ alone.

\begin{center}              
{\bf Acknowledgments} 
\end{center}
We thank Rich Lebed and Nimai Mukhopadhyay for useful conversations.
C.E.C. thanks the National Science Foundation for support under grant 
PHY-96-00415.
\appendix
\section{Appendix}

In the tables below we present the numerical results corresponding
to Figures~1 through 4 in the text, as well as the experimental data.  

\begin{table}[ht]
\begin{tabular}{ccccc}
\\ \hline
Parameters: &  $a_1 = 0.615\pm 0.029$ &  $b_1=-0.294\pm 0.039$ &  
$b_2=-0.297\pm 0.035$ & \\
& $\theta_{N1}=0.597\pm 0.160$ &  $\theta_{N3}=3.060\pm 0.110$ & & \\
\hline\hline
& $A^p_{1/2}$ & $A^p_{3/2}$ & $A^n_{1/2}$ & $A^n_{3/2}$ \\ \hline
$\Delta(1700)$ & 0.084 & 0.083 & - & -  \\  
$\Delta(1620)$ & 0.085 &  - & - & -   \\
$ N(1675) $ & 0 & 0 & -0.035 & -0.050 \\
$ N(1700) $ & -0.002 & 0.012 & 0.015 & -0.019  \\
$ N(1650) $ & 0.047 & - & -0.047 & -  \\ 
$ N(1520) $ & -0.028 & 0.164 & -0.035 & -0.127  \\
$ N(1535) $ & 0.074 & - & -0.074 & -  \\ \hline
\end{tabular}
\caption{Fit displayed in Figure~1, with $\chi^2=52.92$.  The amplitudes
are in units of GeV$^{-1/2}$.  In this fit, the mixing angles were free
to vary.}
\label{table2}
\end{table}

\begin{table}[ht]
\begin{tabular}{ccccc}
\\ \hline
Parameters: &  $a_1 = 0.816\pm 0.061$ &  $b_1=-0.299\pm 0.038$ &  
$b_2=-0.308\pm 0.032$ \\
&$c_3=-0.072\pm 0.020$ & $\theta_{N1}=0.610$ (fixed) &  
$\theta_{N3}=3.04$ (fixed)  \\
\hline\hline
& $A^p_{1/2}$ & $A^p_{3/2}$ & $A^n_{1/2}$ & $A^n_{3/2}$ \\ \hline
$\Delta(1700)$  & 0.070 & 0.057 & - & - \\
$\Delta(1620)$  & 0.063 &  - & - & - \\
$ N(1675) $  & 0 & 0 & -0.036 & -0.051 \\ 
$ N(1700) $ & -0.003 & 0.015 & 0.014 & -0.023 \\ 
$ N(1650) $  & 0.046 & - & -0.058 & - \\
$ N(1520) $  & -0.032 & 0.163 & -0.048 & -0.151 \\
$ N(1535) $  & 0.070 & - & -0.088 & - \\ \hline
\end{tabular}
\caption{Fit displayed in Figure~2, with $\chi^2=39.21$. The amplitudes 
are in units of GeV$^{-1/2}$.}
\label{table3}
\end{table}

\begin{table}[ht]
\begin{tabular}{ccccc}
\\ \hline
Parameters: & $a_1=0.808\pm 0.063$ & $b_1=-0.192\pm0.118$ & 
$b_2=-0.498\pm0.115$ \\
& $c_1=-0.087\pm 0.047$ & $c_2=-0.036\pm 0.044$ & $c_3=-0.077\pm 0.026$ \\
& $c_4=0.049\pm 0.041$ & $d_1=0.077\pm 0.036$ & $d_2=0.009\pm 0.035$ \\
& $d_3=0.013\pm 0.032$ & $d_4=0.012\pm 0.034$ & $\theta_{N1}=0.610$ (fixed) \\
& $\theta_{N3}=3.04$ (fixed) & & \\
\hline\hline
& $A^p_{1/2}$ & $A^p_{3/2}$ & $A^n_{1/2}$ & $A^n_{3/2}$ \\ \hline
$\Delta(1700)$ & 0.077 & 0.061 & - & -   \\  
$\Delta(1620)$ & 0.057 &  - & - & -   \\
$ N(1675) $ & 0.016 & 0.011 & -0.049 & -0.062 \\
$ N(1700) $ & -0.012 & 0.015 & 0.041 &  0.009   \\
$ N(1650) $ & 0.038 & - & -0.071 & -  \\ 
$ N(1520) $ & -0.025 & 0.162 & -0.045 & -0.150  \\
$ N(1535) $ & 0.074 & - & -0.087 & -  \\ \hline
\end{tabular}
\caption{Fit displayed in Figure~3, with $\chi^2=28.87$. The amplitudes
are in units of GeV$^{-1/2}$.}
\label{table4}
\end{table}

\begin{table}[ht]
\begin{tabular}{ccccc}
\\ \hline
Parameters: &  $a_1 = 0.648\pm 0.030$ &  $b_1=-0.278\pm 0.039$ &  
$b_2=-0.294\pm 0.037$ \\
& $\theta_{N1}=0.603\pm 0.158$ &  $\theta_{N3}=3.027\pm 0.143$ & \\ 
\hline\hline
 & $A^p_{1/2}$ & $A^p_{3/2}$ & $A^n_{1/2}$ & $A^n_{3/2}$ \\ \hline
$\Delta(1700)$  & 0.086 & 0.089 & - & - \\
$\Delta(1620)$  & 0.090 &  - & - & - \\
$ N(1675) $  & 0 & 0 & -0.034 & -0.048 \\
$ N(1700) $  & -0.002 & 0.017 & 0.014 & -0.022 \\
$ N(1650) $  & 0.048 & - & -0.050 & - \\
$ N(1520) $   & -0.021 & 0.169 & -0.040 & -0.133 \\
$ N(1535) $  & 0.076 & - & -0.077 & - \\ \hline
\end{tabular}
\caption{Fit displayed in Figure~4, with $\chi^2=22.92$.  The amplitudes
are in units of GeV$^{-1/2}$.  The difference between this fit and that of
Figure~\ref{figure1}/Table~\ref{table2} is the data point for
$A_{1/2}(\Delta(1620)\rightarrow p\gamma)$ was not used in the fit, although
the value resulting from the fit parameters is shown above.}
\label{table5}
\end{table}

\begin{table}[ht]
\begin{tabular}{cccccc}
\\ \hline\hline
&  & $A^p_{1/2}$ & $A^p_{3/2}$ & $A^n_{1/2}$ & $A^n_{3/2}$    \\ \hline
$\Delta(1700)$ &  & $0.104\pm 0.015$ & $0.085\pm 0.022$ & - & -   \\ 
$\Delta(1620)$ &  & $0.027\pm 0.011$&  - & - & -      \\
$ N(1675) $ &  & $0.019\pm 0.008$ & $0.015\pm 0.009$ & $-0.043\pm 0.012$ 
& $-0.058\pm 0.013$   \\
$ N(1700) $ &  & $-0.018\pm 0.013$ & $-0.002\pm 0.024$ & $0.000\pm 0.050$ 
& $-0.003\pm 0.044$   \\
$ N(1650) $ &  & $0.053\pm 0.016$ & - & $-0.015\pm 0.021$ & -  \\ 
$ N(1520) $ &  & $-0.024\pm 0.009$ & $0.166\pm 0.005$ & $-0.059\pm 0.009$ 
& $-0.139\pm 0.011$ \\
$ N(1535) $ &  & $0.070\pm 0.012$ & - & $-0.046\pm 0.027$ & -   \\ \hline
\end{tabular}
\caption{Experimental data, from Ref.~\protect\cite{rpp}, in units
of GeV$^{-1/2}$.}
\label{table6}
\end{table}


\begin{thebibliography}{99} 
\frenchspacing
\bibitem{cgo}
C. Carone, H. Georgi, and S. Osofsky, Phys. Lett. {\bf B322}, 227 (1994).
\bibitem{ml}
M. A. Luty and J. March-Russell, Nucl. Phys. {\bf B426}, 71 (1994);
M. A. Luty, J. March-Russell, M. White, Phys. Rev. D {\bf 51}, 2332 (1995). 
\bibitem{djm1}
R. Dashen and A. Manohar, Phys. Lett. {\bf B315}, 425 (1993); {\bf 315},
438 (1993); E. Jenkins, {\em ibid.} {\bf 315}, 431 (1993); {\bf 315}, 441
(1993); {\bf 315}, 447 (1993); R. Dashen, E. Jenkins, and A.V. Manohar, 
Phys. Rev. D {\bf 49} 4713 (1994).
\bibitem{djm2}
R. Dashen, E. Jenkins, and A.V. Manohar, Phys. Rev. D {\bf 51}, 3697 
(1995).
\bibitem{cgkm}
C. D. Carone, H. Georgi, L. Kaplan, and D. Morin, Phys. Rev. D {\bf 50}, 5793
(1994).
\bibitem{py}
D. Pirjol and T-M Yan, Phys. Rev. D {\bf 57}, 1449 (1998);
CLNS-97-1520, Nov. 1997, hep-ph/9711201.
\bibitem{close}
F.E. Close, {\em Quarks and Partons}, Academic Press, London, 1979.
\bibitem{witten}
E. Witten, Nucl. Phys. {\bf B160}, 57 (1979).
\bibitem{rpp}
Review of Particle Properties, R.M. Barnett, C.D. Carone, {\em et al.} (The 
Particle Data Group), Phys. Rev. D {\bf 54}, 1 (1996).
\bibitem{hey}
A.J.G. Hey, P.J. Litchfield, and R.J. Cashmore, Nucl. Phys. {\bf B95}, 516 
(1975).
\end{thebibliography}
\end{document}